\documentclass[sigconf]{acmart}

\newif\ifDEBUG
\DEBUGtrue

\newif\ifEXTENDED
\EXTENDEDfalse

\usepackage{graphicx}
\usepackage{hyperref} 
\usepackage{cleveref}
\usepackage{pifont}
\usepackage{xcolor}
\usepackage{listings}
\usepackage{minted}
\usepackage{subcaption}
\usepackage{caption}

\usepackage{booktabs}
\usepackage{multirow} 
\usepackage{subcaption}
\usepackage{makecell}
\usepackage{tabularx}
\usepackage{ragged2e}
\usepackage{xspace}
\usepackage{soul}
\usepackage{array} 
\usepackage{float}

\usepackage{amsmath}
\usepackage{textcomp}
\usepackage{colortbl}
\usepackage{longtable}
\usepackage{adjustbox}
\usepackage{algorithm}
\usepackage{threeparttable}
\usepackage{tablefootnote}
\usepackage{booktabs} 
\usepackage{amsmath}
\usepackage{algorithm}
\usepackage{algpseudocode}
\usepackage[normalem]{ulem} 
\usepackage{xspace}
\usepackage{relsize}
\usepackage{multirow} 
\usepackage{balance} 
\usepackage{fancyvrb} 
\usepackage{pifont}

\usepackage{enumitem}
\setlist[itemize]{leftmargin=*,noitemsep,topsep=0pt}
\setlist[enumerate]{leftmargin=*}

\usepackage{etoolbox}
\makeatletter
\patchcmd{\@makecaption}
	{\scshape}
	{}
	{}
	{}
\makeatletter
\patchcmd{\@makecaption}
	{\\}
	{.\ }
	{}
	{}
\makeatother

\newcommand{\eg}{\textit{e.g.,}\xspace}

\usepackage{mdframed}
\mdfsetup{skipabove=0.5\topskip,skipbelow=0.5\topskip,align=center}

\usepackage{amsthm}
\newtheorem{thm}{Theorem}

\DeclareMathSymbol{\mlq}{\mathord}{operators}{``}
\DeclareMathSymbol{\mrq}{\mathord}{operators}{`'}

\newif\ifSAVESPACE
\SAVESPACEfalse

\ifSAVESPACE

\else

\fi

\usepackage{todonotes}
\usepackage{soul}
\usepackage{fontawesome}
\usepackage{ragged2e}

\ifDEBUG
    \newcommand{\AH}[1]{\todo[color=cyan,inline]{AH:#1}}
    \newcommand{\AM}[1]{\todo[color=red,inline]{Machiry:#1}}
    \newcommand{\JD}[1]{\todo[color=yellow,inline]{JD:#1}}
    \newcommand{\SA}[1]{\todo[color=green,inline]{SA:#1}}
    \newcommand{\PA}[1]{\todo[color=orange,inline]{PA:#1}}
    
    \newcommand{\KR}[1]{\todo[color=yellow,inline]{Kyle:#1}}
    \newcommand{\LS}[1]{\todo[color=green,inline]{LS:#1}}
    \newcommand{\HP}[1]{\todo[color=cyan,inline]{HP:#1}}
    \newcommand{\NJE}[1]{\todo[color=red,inline]{NJE: #1}}
    \newcommand{\GKT}[1]{\todo[color=red,inline]{GKT:#1}}
    
    \newcommand{\RH}[1]{\todo[color=red,inline]{RH:#1}}
    \newcommand{\WJ}[1]{\todo[color=SkyBlue,inline]{Wenxin:#1}} 
    \newcommand{\KC}[1]{\todo[color=orange,inline]{Kelechi Says:#1}}
    \newcommand{\AG}[1]{\todo[color=orange,inline]{AG:#1}}
    \newcommand{\PJ}[1]{\todo[color=lime,inline]{PJ:#1}}
    \newcommand{\AZ}[1]{\todo[color=teal,inline]{Antonio:#1}}
    \newcommand{\PT}[1]{\todo[color=pink,inline]{Parth:#1}}
    
\else
    \newcommand{\AH}[1]{}
    \newcommand{\AM}[1]{}
    \newcommand{\JD}[1]{}
    \newcommand{\SA}[1]{}
    \newcommand{\PA}[1]{}
    \newcommand{\KR}[1]{}
    \newcommand{\LS}[1]{}
    \newcommand{\HP}[1]{}
    \newcommand{\NJE}[1]{}
    \newcommand{\GKT}[1]{}
    \newcommand{\KC}[1]{}
    \newcommand{\RH}[1]{}
    \newcommand{\WJ}[1]{}
    \newcommand{\AG}[1]{}
    \newcommand{\PJ}[1]{}
    \newcommand{\PT}[1]{}
    \newcommand{\AZ}[1]{}
    
\fi

\usepackage{hyperref}

\usepackage{cleveref}
\crefformat{section}{\S#2#1#3}
\crefname{figure}{Figure}{Figures}
\crefname{table}{Table}{Tables}
\crefname{theorem}{Theorem}{Theorems}
\crefname{thm}{Theorem}{Theorems}
\crefname{lemma}{Lemma}{Lemmata}
\crefname{equation}{Eqt.}{Eqts.}
\crefformat{Grammar}{Grammar #1}
\crefname{appendix}{Appendix}{Appendices}
\crefname{listing}{Listing}{Listings}

\newcommand{\myparagraph}[1]{\paragraph*{\textbf{#1}}}
\renewcommand{\myparagraph}[1]{\vspace{0.25em} \hspace{0.1cm}\underline{\textit{#1:}}}

\usepackage{url}

\usepackage{tcolorbox}

\makeatletter
\newcommand{\linebreakand}{%
  \end{@IEEEauthorhalign}
  \hfill\mbox{}\par
  \mbox{}\hfill\begin{@IEEEauthorhalign}
}
\makeatother

\usepackage{pifont}

 \usepackage{pifont}

\begin{document}

\title{Beyond Local Code Optimization: Multi-Agent Reasoning for Software System Optimization}






\author{Huiyun Peng$^{*}$, Parth Vinod Patil$^{\dagger}$, Antonio Zhong Qiu$^{*}$, George K. Thiruvathukal $^{\ddagger}$, James C. Davis$^{*}$}
\affiliation{%
  \institution{$^{*}$Purdue University, $^{\dagger}$Amazon Robotics, $^{\ddagger}$Loyola University Chicago}
  \country{USA}
}

\renewcommand{\shortauthors}{Peng et al.}

\begin{abstract}
Large language models and AI agents have recently shown promise in automating software performance optimization, but existing approaches predominantly rely on local, syntax-driven code transformations. 
This limits their ability to reason about program behavior and capture whole system performance interactions. 
As modern software increasingly comprises interacting components --- such as microservices, databases, and shared infrastructure --- effective code optimization requires reasoning about program structure and system architecture beyond individual functions or files.

This paper explores the feasibility of whole system optimization for microservices. We introduce a multi-agent framework that integrates control-flow and data-flow representations with architectural and cross-component dependency signals to support system-level performance reasoning. The proposed system is decomposed into coordinated agent roles—summarization, analysis, optimization, and verification—that collaboratively identify cross-cutting bottlenecks and construct multi-step optimization strategies spanning the software stack.
We present a proof-of-concept on a microservice-based system that illustrates the effectiveness of our proposed framework, achieving a 36.58\% improvement in throughput and a 27.81\% reduction in average response time.
\end{abstract}



\maketitle

\section{Introduction}
Large Language Models (LLMs) are transforming modern software engineering, demonstrating strong performance in tasks such as code generation, refactoring, program repair, and documentation~\cite{Hou2024}. 
More recently, LLMs and agent-based systems have been explored for code optimization, where models propose code and configuration changes aimed at improving runtime efficiency, maintainability, or resource utilization~\cite{peng2024largelanguagemodelsenergyefficient, luo2025largelanguagemodelagent, shypula2024learning}. 
These advances suggest the potential for AI systems to assist performance engineering, a traditionally labor-intensive and expertise-driven activity~\cite{Balaprakash2018}.



Despite this progress, existing LLM-based optimization methods predominantly operate at local program scopes, such as individual functions or classes~\cite{peng2025sysllmaticlargelanguagemodels, gong2025}. 
While effective for localized optimizations, this focus overlooks the fact that real-world performance bottlenecks often emerge from interactions across multiple components and layers of a software system, such as databases, middleware, and shared infrastructure~\cite{smith2000}. 
As a result, purely function- or file-level reasoning struggles to capture whole system behavior or identify cross-cutting bottlenecks. 
Addressing these challenges requires reasoning over richer program and architectural abstractions that expose control and data flow, and component interactions.

In this work, we propose an agentic framework for software optimization that elevates performance reasoning from local code edits to whole software systems. 
Our framework uses static analysis to extract control- and data-flow evidence within components and interaction structure across components, such as service boundaries and call graphs, and aggregates these signals into performance-oriented software summaries. 
To scale this reasoning to complex systems, we decompose the framework into coordinated agent roles --- system summarization, bottleneck analysis, optimization, and verification --- that collaborate to construct multi-step optimization plans across the software stack.


We show a proof-of-concept of our prototype on TeaStore~\cite{Teastore}, a  Java-based microservices application, to demonstrate its effectiveness.
The results show that our prototype achieves a 36.58\% increase in throughput and a 27.81\% reduction in average response time, while preserving correctness, highlighting promising directions for AI-driven whole system performance engineering.

\ul{In summary, we contribute}:
\begin{itemize}
    \item \textbf{A Novel Framework (\cref{sec:design}):} We introduce a multi-agent optimization framework that integrates static analysis to reason about cross-component interactions and performance bottlenecks.
    \item \textbf{Proof of Concept (\cref{sec:prototype_and_result}):}
    We evaluate the system on a representative microservice to demonstrate its effectiveness.
\end{itemize}
\vspace{0.15em}
We submit this work to JAWS to obtain early feedback on the problem framing and system design. We hope to incorporate the feedback and submit an extended version to TSE or TOSEM.

\begin{figure*}[t]
    \centering
    \includegraphics[width=\linewidth, trim={1.4cm 11.9cm 1.2cm 2cm}, clip]{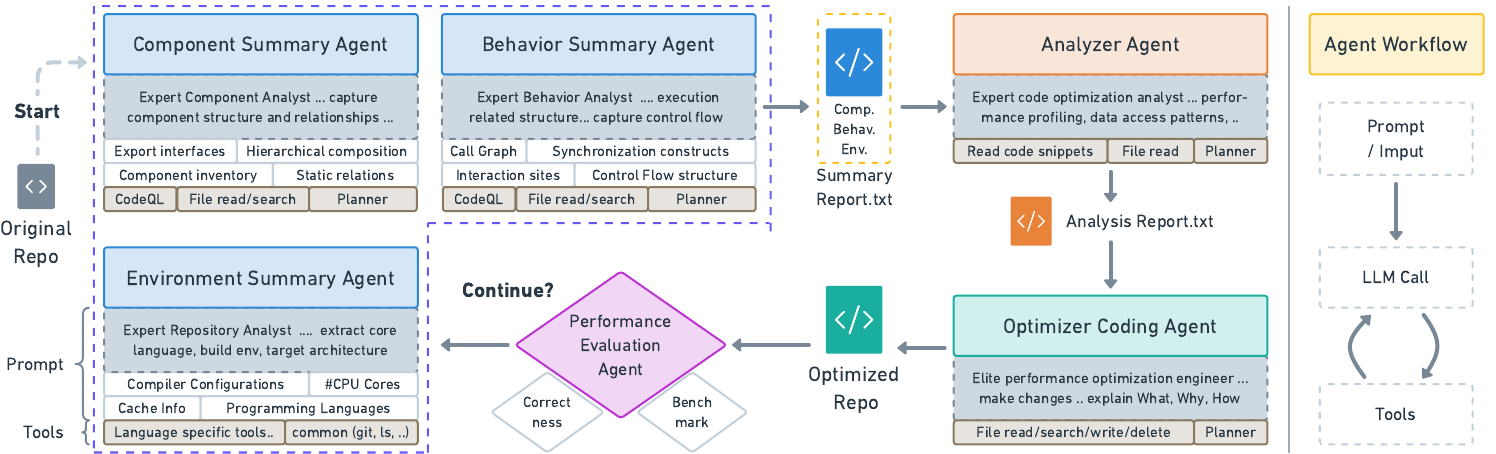}
    \caption{
        Design overview.
        The system cycles through summarization, analysis, optimization, and evaluation.
        It terminates when no further optimizations can be applied (as determined by the performance evaluation agent).
    }
    \vspace{-1em}
    \label{fig:methodology-overview}
\end{figure*}

\section{Background}
\paragraph*{\textbf{Program Analysis}}
Static program analysis enables reasoning about program behavior without execution by modeling control flow, data flow, and interprocedural dependencies~\cite{Aho1986,muchnick1997advanced}.
Core representations such as control-flow graphs and data-flow analyses capture execution structure and value propagation, forming the foundation for compiler optimization, performance modeling, and bug detection~\cite{Allen1970Control,Kam1976GlobalDF,Tip1994ASO}.
Beyond local code properties, static analysis has been applied to expose performance-relevant behavior arising from interactions across components, libraries, and data-access layers, demonstrating its effectiveness in identifying system-level inefficiencies~\cite{chen2014,turcotte2023, Chen2016Cache,chen2016,lyu2018,nagy2018}.
\textit{These capabilities motivate the use of static analysis as a foundational abstraction for guiding automated performance reasoning in our framework.}

\paragraph*{\textbf{Software Architecture}}
Software architecture captures the high-level structure of a system, describing its components, their interactions, and the design decisions that shape system-wide qualities such as performance and scalability~\cite{perry1992foundations,garlan1993introduction}.
In modern architectures, particularly microservices, performance bottlenecks often arise from cross-component dependencies, communication patterns, and request-routing paths rather than isolated code segments~\cite{dragoni2017microservices}.
Prior work shows that architectural abstractions enable reasoning about such performance effects and support performance-improving design and refactoring decisions~\cite{smith2000performance,smith2001software,Becker2007,Petriu2000,Avritzer2025}.
\textit{Building on these insights, our work leverages architectural abstractions to enable agentic performance optimization beyond local-code reasoning.}

\section{Problem Statement}
Modern software systems comprise interacting components whose performance behavior emerges from execution structure, architectural organization, and cross-component interactions~\cite{martin2017cleanarchitecture, Kruchten1995}. 
Many performance issues stem not from isolated code inefficiencies but from architecture- and design-level decisions that shape component interaction and resource usage~\cite{smith2000}. Such system-level problems are often indirect and context-dependent, making them difficult to detect through local inspection and consequently under-addressed in practice~\cite{Zhao2020}. Improving system performance while preserving correctness remains a fundamental challenge.

Performance analysis and optimization of software systems has been studied for decades across modeling, analytical, and architectural approaches, but many of these techniques require substantial manual effort and remain difficult to automate and scale~\cite{Jain1991, balaprakash2018autotuning}.
Recent LLM–based techniques have shown promise in automated code optimization due to their ability to handle many programming languages and real-world implementations.
However, existing approaches primarily operate at localized program scopes, leaving it unclear whether automated techniques can reason about performance behavior that emerges from cross-component and architectural interactions~\cite{gong2025, peng2025sysllmaticlargelanguagemodels}.
This begs the question:
\emph{Can automated techniques identify and reason about whole system performance issues arising from architectural structure and cross-component interactions, beyond isolated code scopes, while preserving correctness?}

\section{Design}
\label{sec:design}
We propose a multi-agent framework for whole system optimization, as shown in \cref{fig:methodology-overview}. 
The framework decomposes performance engineering into a four-stage pipeline: (1) a \textit{Summarization Agent} that extracts system structure and runtime characteristics; (2) an \textit{Analysis Agent} that identifies performance bottlenecks; (3) an \textit{Optimization Agent} that generates targeted code improvements; and (4) a \textit{Performance Evaluation Agent} that measures optimization impact. 
The framework is stateful: agents operate over persistent representations produced by earlier stages, preserving architectural relationships, cross-component dependencies, and optimization context to enable whole system reasoning.

\subsection{Summarization Agent}
\label{sec:summarization-agent}

\begin{table}[h]
\centering
\caption{Information captured by Component Summary and Behavior Summary agents using static analysis.}
\label{tab:agent-summaries}
\vspace{-1em}

\small
\begin{tabular}{@{}p{0.38\linewidth}p{0.56\linewidth}@{}}
\toprule
\textbf{Captured Information} & \textbf{Description} \\
\midrule
\multicolumn{2}{@{}l}{\textbf{\textit{Component Summary Agent}}} \\
Component inventory & Identifies Functions, classes, packages, and services from code structure. \\
Hierarchical composition & Derives Parent-child relationships (service $\to$ package $\to$ class $\to$ method). \\
Exported interfaces & HTTP endpoints, and API entry points. \\
Static dependencies & Extracts Call-based, type-based, and resource-based component dependencies. \\
Evidence mapping & Associate with precise source file locations \\
\midrule
\multicolumn{2}{@{}l}{\textbf{\textit{Behavior Summary Agent}}} \\
Inter-procedural call graphs & Construct Static call graphs rooted at entry points, capturing component  invocations\\
Control-flow structure & Identify control flow with components \\
Interaction sites & Service calls, DB access, external relations. \\
Synchronization constructs & Identifies synchronization points (locks). \\
Static evidence linkage & Behavior abstractions linked to analysis evidence for reproducibility. \\
\bottomrule
\end{tabular}
\vspace{-1em}
\end{table}

Motivated by architectural modeling distinctions between structural and behavioral views of software systems~\cite{Zhao2024}, we decompose code summarization into two specialized agents: a \emph{component summary agent} and a \emph{behavior summary agent} (\cref{tab:agent-summaries}). We additionally introduce an \emph{environment summary agent} to capture execution and deployment context that is not intrinsic to the code but is necessary for software system performance reasoning~\cite{Meng2017}.

The \emph{Component Summarization Agent} extracts the structural architecture of the software system. 
It identifies service boundaries and architectural components, and classifies them by architectural role (\eg service endpoints, data access, and integration interfaces).
The agent records hierarchical and compositional relationships among components, documents exposed interfaces including service endpoints and internal APIs, and constructs dependency mappings that capture call-based and resource-based coupling within and across services. 
The resulting representation encodes data such as service topology and inter-component dependency structure.

The \emph{Behavior Summarization Agent} captures execution-related properties that influence system performance. It analyzes interprocedural control flow to model execution paths across components, summarizes control-flow complexity, and identifies interactions with databases. The agent also characterizes concurrency by detecting synchronization constructs and thread-safety mechanisms.

The \emph{Environment Summary Agent} captures runtime and deployment context extracted from build configurations and artifacts. It extracts information about programming languages, compilers, configuration parameters, dependencies, and project-level settings from build files and deployment artifacts. By making environmental assumptions explicit, this agent ensures that downstream analysis and optimization account for configuration- and infrastructure-level constraints in addition to code-level structure and behavior.

\subsection{Analysis Agent}
\label{sec:analysis-agent}
The \emph{analysis agent} identifies and prioritizes performance optimization opportunities using static analysis summaries. 
It consumes structured representations of the system, such as call graphs, hotspots, dependency information, and static signals such as database access patterns and service interaction topologies. 
The agent employs a multi-stage reasoning process: (1) extract performance signals from summary data, (2) inspect performance-critical code regions, and (3) rank optimization opportunities by estimated impact and confidence.
The output is a structured report that specifies evidence-backed priorities with source locations and impact estimates, and highlights candidate files, risks, and gaps for further investigation.

\subsection{Optimization Agent}
\label{sec:optimization-agent}

The \emph{optimization agent} translates high-level optimization intents into concrete, verifiable system modifications. Guided by the analysis agent’s findings, it applies targeted code changes that correspond to the identified optimization opportunities. 
To preserve system correctness and compatibility, all modifications are constrained to be non-breaking, disallowing changes to public APIs or service interfaces. The agent produces a structured optimization report that documents each applied change using unified diffs, accompanied by technical justifications that explain the expected performance impact and its alignment with the analyzed system context. 

\subsection{Performance Evaluation Agent}
\label{sec:performance-evaluation-agent}

The \emph{performance evaluation agent} validates both functional correctness and performance impact.
We assume the target software system is accompanied by a reliable automated test suite.
Each candidate optimization is first subjected to test-based correctness checking; optimizations that fail any test are rejected.
For valid candidates, the agent performs dynamic profiling under representative workloads and compares end-to-end metrics, including latency, throughput, and resource utilization, against baseline measurements.
The evaluation outcomes are fed back to guide subsequent optimization decisions, enabling iterative refinement toward performance improvements while preserving correctness.

\section{ Prototype and Preliminary Results}
\label{sec:prototype_and_result}
This section presents a prototype of the proposed framework and reports proof-of-concept results demonstrating its feasibility.
\subsection{Prototype Implementation}
We implement a prototype of the proposed agentic optimization framework (Summarization, Analysis, and Optimization) in Python. The system comprises 29 Python source files and a total of 8,000 lines of code. 
In the current prototype, correctness validation and performance evaluation are conducted separately from the system, with full integration into the automated pipeline left to future work.

\paragraph*{Agent}
We use \textsc{LangGraph} to implement the pipeline as a directed graph of modular agents (summarization, analysis, optimization, and evaluation) with explicit control flow and shared state. 
Agents consume structured inputs and pass outputs to downstream agents, while shared state preserves contextual information such as summaries, identified performance issues, and applied optimizations. 
For analysis and reproducibility, we integrate \textsc{LangSmith} to track model outputs, intermediate states, and execution traces.

\paragraph*{Static Analysis Tool}
We leverage \textsc{CodeQL} as the static analysis engine for extracting semantic program abstractions. \textsc{CodeQL} models source code as a queryable database over abstract syntax trees, control flow graphs, and call graphs. We develop language-specific query packs that extract component structure, architectural relationships, and execution patterns, serializing them into a language-agnostic representation for the multi-agent pipeline.

\subsection{Preliminary Results}
\label{sec:result}
\subsubsection{Experimental Setup}
We use \textsc{TeaStore}~\cite{Teastore} to illustrate the effectiveness of the proposed system. \textsc{TeaStore} is a Java-based microservice that models an online retail application composed of multiple interacting services connected via REST APIs. It contains six services that implement end-to-end workflows.
Performance is affected by cross-service request paths and shared resources.

We use \texttt{gpt-5.2} (SOTA in Jan. 2026) as the LLM for all agents in the pipeline. Following prior work~\cite{garg2025rapgenapproachfixingcode, shypula2024learning, Gao2025}, we configure the LLM inference temperature to 0.7 to balance exploration and determinism. We conduct all evaluations on a dedicated bare-metal server (Intel Xeon W-2295, 36 CPUs, 188 GB RAM). 
Correctness is validated using the JUnit test suites provided by \textsc{TeaStore}.
We benchmark both implementations using \textit{Apache JMeter v5.6.3}~\cite{jmeter} with the standardized load-testing configuration provided by \textsc{TeaStore}.
Each workload executes 800,000 requests under identical conditions, and end-to-end performance metrics are collected.

\subsubsection{Preliminary Results}

Table~\ref{tab:optimizations_and_results} illustrates the static analysis signals and corresponding optimizations identified by our system: (O1) HTTP client reuse via singleton patterns to avoid redundant connection pool initialization, (O2) lock contention removal by replacing synchronized methods with volatile flags on request-critical paths, and (O3) reduced object allocation through shared \texttt{ObjectMapper} instances.
End-to-end performance benchmarking on \textsc{TeaStore} shows consistent improvements in throughput, response time, and latency, demonstrating the potential of agent-based frameworks for effective whole-system optimization.
The full optimization run consumes approximately 1.6M tokens, costs \$1.28, and takes roughly 24 minutes.

\begin{table}[h]
\centering
\caption{
Representative optimizations identified by our framework and end-to-end performance improvements.
}
\label{tab:optimizations_and_results}
\vspace{-0.5em}

\small
\begin{tabular}{p{0.25\linewidth}p{0.32\linewidth}p{0.30\linewidth}}
\toprule
\textbf{Static Evidence} & \textbf{Interpretation} & \textbf{Suggestion} \\
\midrule
\textbf{O1: HTTP Client} & & \\
Jersey \texttt{Client} created per \texttt{RESTClient} constructor. Expensive initialization on service call paths.
& 
Duplicate connection pools across microservices increase resource overhead, degrading throughput and increasing tail latency on service-to-service calls.
& 
Singleton pattern using \texttt{SHARED\_CLIENT} \texttt{\_HTTP/S} for thread-safe reuse with connection pooling benefits.
\\
\midrule
\textbf{O2: Volatile Flag} & & \\
\texttt{synchronized} on \texttt{setMaintenance} \texttt{-ModeInternal()} (CodeQL analysis).
& 
Unnecessary lock contention on request-critical paths introduces thread blocking, causing latency spikes under concurrent access.
& 
Replace synchronized method with \texttt{volatile boolean} flag for lock-free access while ensuring memory visibility.
\\
\midrule
\textbf{O3: ObjMapper} & & \\
\texttt{new ObjectMapper()} per request in session methods of \texttt{AbstractUI} \texttt{-Servlet} class.
& 
Per-request allocation in the base servlet propagates across all UI request paths, increasing GC pressure and CPU serialization overhead.
& 
Shared \texttt{static final} instance that remains thread-safe after configuration to eliminate allocations.
\\
\bottomrule
\end{tabular}

\vspace{0.25cm}

\begin{tabular}{lrrr}
\toprule
\textbf{Metric} & \textbf{Original} & \textbf{Optimized} & \textbf{Improvement} \\
\midrule
Throughput (req/sec) & 1197.79 & 1635.89 & \textbf{+36.58\%} \\
Avg Response Time (ms) & 12.84 & 9.27 & \textbf{$-$27.81\%} \\
P50 Latency (ms) & 13.00 & 9.00 & \textbf{$-$30.77\%} \\
P90 Latency (ms) & 23.00 & 18.00 & \textbf{$-$21.74\%} \\
P99 Latency (ms) & 26.00 & 23.00 & \textbf{$-$11.54\%} \\
\bottomrule
\end{tabular}
\vspace{-1em}
\end{table}

\section{Future Plans}

\myparagraph{Extensions to the System} This work presents a prototype of an agentic system for software optimization.
Future work will extend the prototype with specialized agents for performance impact prediction and optimization validation across workloads and environments, and strengthen agent coordination through richer intermediate representations and iterative feedback.
We will also support measurement of individual optimization patch impact, enabling cost-aware prioritization of patches that deliver the greatest performance gains relative to optimization cost.

\myparagraph{Metrics and Measurement}
We will evaluate optimization correctness and effectiveness along three dimensions.
\emph{Correctness:} All generated changes will be validated using existing test suites and manual inspection to ensure functional correctness.
\emph{Performance impact:} We will measure optimization impact using performance metrics such as latency, throughput, and resource utilization, as supported by the evaluated workloads.
\emph{System cost:} We will quantify the cost of our framework using total token consumption, end-to-end execution time, and monetary cost, to assess practical deployability.

\myparagraph{Ablations}
We will ablate along three dimensions.
\begin{enumerate}
    \item \textit{Summarization ablation:} We will selectively remove individual summarization signals, including component-level, behavioral-level, and environment-level context, to assess the contribution of each abstraction to downstream performance analysis.
    \item \textit{Agentic component ablation:} We will construct reduced system variants that selectively remove the summarization agent, the analysis agent, or both. This ablation isolates the impact of structured architectural understanding and analytical reasoning on downstream optimization effectiveness.
    \item \textit{System-wide ablation:} We will vary the configuration of the agent interaction loop, including the number of optimization iterations and feedback passes, to study the role of iterative refinement in optimization effectiveness and stability.
\end{enumerate}

\myparagraph{Benchmark and Baseline Comparison}
We will evaluate our framework across a diverse set of software systems with varying architectures, scales, performance characteristics, and languages (\eg DeathStarBench~\cite{deathstarbench}), to assess generalizability beyond a single application. 
We compare against two baseline categories:
\begin{enumerate}
    \item \textit{Repository-level optimization techniques:} We will compare against SysLLMatic~\cite{peng2025sysllmaticlargelanguagemodels}, a state-of-the-art LLM-based system for profiling-guided, repository-level code optimization.
    \item \textit{Agent-based optimization systems:} We will evaluate against recent agentic software engineering approaches such as OpenCode~\cite{OpenCodeAI2026} and CodeX~\cite{OpenAICodexGitHub2026}, which employ multi-agent workflows for automated code modification.
\end{enumerate}
While non-LLM approaches (\eg traditional compiler optimization and autotuning) are widely used, we do not include them as baselines because our system leverages compiler configuration as part of the environment summarization and optimization context.

\myparagraph{Open-Source Models}
To improve reproducibility, we will also evaluate the system using open-source LLMs of varying sizes and architectures, assessing whether the proposed framework remains effective independent of specific model capabilities.

\vspace{0.5em}
\noindent
\textit{Data Availability:}
\label{sec:DataAvailability}
Our script, prompt, and data are available at~\cite{artifact}.

\bibliographystyle{ACM-Reference-Format}
\bibliography{references/reference}

\end{document}